\documentstyle[12pt]{article}
\def\frac#1#2{{\displaystyle#1\over\displaystyle#2}}
\begin{document}

\baselineskip=4ex

\title{KINETIC THEORY OF ANOMALOUS TRANSPORT OF SUPRATHERMAL PARTICLES.}

\author{A.V. Gurevich, A.V. Lukyanov and K.P. Zybin.}

\date{}
\maketitle

\begin{abstract}
Investigation of behavior of fast electrons in toroidal discharges
was performed. The kinetic equation, describing evolution
of fast particle distribution was derived and analyzed. Semi - analytical
solution of the kinetic equation was obtained into suprathermal energy region.
External electric field, anomalous transport, nonuniformity of mean magnetic
field and collisions was shown to be an important factors, affecting
distribution function. The role of ambipolar electric field was established and
identified as an essential factor of the process of diffusion  of fast
electrons. The effect of strong influence of density profile on diffusion of
fast particles was clearly demonstrated.  Comparison with the experimental
data,
obtained on ZT -40M device, was carried out.  The agreements with the results
of
these experiments is observed.
\end{abstract}

\section{Introduction}

Plasma confinement in toroidal devices is an important problem of thermonuclear
fusion research.  It is well known that various instabilities leading to
discharge turbulization are readily excited in plasma.  Experiments have shown
that electron heat conductivity is by several orders of magnitude higher than
the limit predicted by the neoclassical theory \cite{1} .  Such a strong energy
transfer to a discharge wall is the main energy loss channel in toroidal
systems.  This is the reason why this problem is of great interest, see the
review  \cite{2}.

The theory suggests two possible mechanisms of
anomalous transport.  One of these mechanisms is due to turbulence
generated by potential electric field oscillations caused by drift
instabilities  \cite{3,4,5,6}.  The other mechanism is due to
turbulence induced by magnetic fluctuations. On exceeding the threshold value
(determined by overlapping of mean resonance modes), magnetic field
fluctuations lead to a stochastic wandering of magnetic field lines
about the discharge and, as a consequence, to a transport of particles
moving along these field lines.  The qualitative picture of this
phenomenon was first presented in the paper  \cite{8}  as a stochastic
description of a diffusion of magnetic field lines.  The theory of this
mechanism of anomalous transport was further on comprehensively studied
by many authors  \cite{9,10,11,12,12a}.  According to the
theory,  the effective diffusion coefficient in the former mechanism is
inversely proportional to the velocity of particle motion

\begin{equation} \label{eq:1}
D_{\perp e}\sim \frac{\langle e^{2}\rangle}{ B^{2}_{0}}
\frac{c^{2}L_{c}}{v_{e}} \end{equation}

\noindent
whereas in the latter case the particle diffusion coefficient increases with
increasing particle velocity

\begin{equation}
\label{eq:2}
D_{\perp b}\sim\frac{\langle b^{2} \rangle }{ B^{2}_{0}}  \, L_{c}v_{e}
\end{equation}

\noindent
The role of each of these mechanisms in the formation of energy flux has not
yet
been finally established.  Measurements of the excitation level in
tokamaks have shown that the first type transport mechanism dominates in
the vicinity of discharge boundary  \cite{13,13a,13aa,13aaa}, while inside a
discharge (where the level of magnetic fluctuations is difficult to measure)
the transport may be due to both excitation mechanisms.  As distinct from the
tokamak,  in pinch type devices with a reversed field, where the magnetic
fluctuation level is by two orders of magnitude greater than that in tokamaks
\cite{14}, predominant is a transport due to large-scale magnetic fluctuations
\cite{15}.

Along with a thermal particle flux, there
also exists a superthermal fast particle flux.  Moreover,  if the
particle lifetime in a discharge increases, the number of superthermal
particles may increase appreciably.  This effect is obviously of
particular importance in devices in which plasma is heated in an ohmic
way.  One of such devices is a reversed field pinch  (RFP)  where,
according to experimental data,  a considerable part of energy is
transported by fast particles \cite{16,17,18}.

To estimate the
energy confinement efficiency in  RFP  type devices,  it is necessary to
determine the distribution function  of fast particles in a
discharge.  Furthermore,  fast particles bear information on the
processes proceeding inside a discharge, and so the investigation of
their distribution may be used to diagnose the state of the plasma
\cite{16,19}.

In the paper  \cite{19} , the authors formulated a consistent
kinetic theory of anomalous transport processes in a turbulized plasma and
derived the kinetic equation describing the averaged particle distribution
function in
these conditions.  A relaxed state of a turbulent plasma and the
anomalous transport processes under  RFP  conditions were analyzed in
the papers \cite{21,22}.  Here we are considering the
superthermal electron distribution function in  RFP.  As has already been
mentioned above, predominant in  RFP  is an anomalous transport due to magnetic
fluctuations.  Therefore, the influence of this particular mechanism of
transport upon the distribution function of superthermal electrons will be our
prime concern in the sequel.  The anomalous transport produces a strong effect
on fast particles because the diffusion coefficient  (\ref{eq:2})  grows
linearly with increasing particle velocity.  Another factor affecting strongly
the distribution function of electrons is an electric field applied to the
plasma.

It is
known that even a weak electric field applied to a plasma induces the
formation of a tail of runaway electrons in the energy range exceeding
the critical one, $\epsilon > \epsilon_{c}$,  where
$\epsilon_{c}=\frac{E_{c}}{E} \, T_{e}$, and  $E_c$  is the critical field
\cite{gr1,23a}.  In  RFP,  an applied field is high, that is, the ratio  $E /
E_c$  is significantly larger than that in tokamaks, and therefore the
distribution function distortion due to the action of the field  $E$  is much
stronger here.  In addition to the vortex electric field, applied to plasma,  a
potential electric field, a so-called "ambipolar field is generated in the
discharge  due to the difference in the ion and electron coefficients
of diffusion.  Producing an immediate effect upon fast electron
diffusion, this field is responsible for the dependence between the ejection of
fast electrons and the transport of the ion plasma component, which have
typically been considered independently.  On the other hand,  as the number of
fast particles increases, they themselves may start affecting the macroscopic
state of the plasma, which is for example the case with convective transfer in
a
rippled field  \cite{24}.  We shall not consider here the effects due to
corrugation, but this role will be played by anomalous transport in a turbulent
plasma.  Thus, an examination of the distribution function of electrons is
necessary for a correct self-consistent analysis of a macroscopic relaxed
state.

Besides the factors listed above, the distribution function is noticeably
affected by inhomogeneity of the mean magnetic field and temperature. For
example, nonuniformity of electron temperature results in thermal
runaway and appearance of hot particle tail in the cold plasma region
\cite{26}.

It should be noted that the influence of anomalous transport in
RFP  on the distribution function of fast electrons was examined in the paper
\cite{27}, but
the authors proceeded from the model kinetic equation which disregards a
number of essential factors affecting the distribution function  In
particular,  the effect of Coulomb collisions, a potential ambipolar
field and inhomogeneity of the mean magnetic field were neglected.  It
is therefore necessary to investigate the distribution function of superthermal
electrons
more thoroughly making allowance for the influence of all essential
factors.  This is just the goal of the present paper.  In section 2 we
derive the master kinetic equation with account of the influence of the
applied electric field, an anomalous diffusion, a potential ambipolar
field, collisions and magnetic field inhomogeneity.  The principal
parameters determining the distribution function of fast electrons are
discussed.
Bearing in mind complicity of studying the complete problem, at the
beginning of section 3 we analyze the distribution function of fast electrons
in a
homogeneous magnetic field.  The strong influence of inhomogeneity of
the electric field applied to plasma and the dependence of the electron
diffusion rate on the profile of the thermal particle density are
resolved.  In section 4 we investigate the influence of inhomogeneity of
the mean magnetic field upon the distribution function of electrons and point
out a
substantial deformation of this function.  Finally, in section 5 we
estimate the influence of the indicated effects in specific  RFP
conditions and compare the developed theory with available experimental
data.  The results of experiments are seen to be in
agreement with the theory.

\section  {The kinetic equation for superthermal
     electrons}

Let us consider a magnetized plasma with a magnetic field $\vec{B}(\vec{r})$
which has a regular
$\vec{ B}_0(\vec {r})$
and a fluctuational  $\vec {b}$
components.  The amplitude of fluctuations will be assumed small as compared
with the mean field $\vec{B_{0}}$, $ |\vec{b}| \ll |\vec{B_{0}}| $.  The basic
quantities characterizing the fluctuations  (the correlation length and the
correlation time)  will be thought of as large as compared to the Larmor radius
of particles and their inverse gyrofrequencies as in \cite{19}.  Within such a
statement of the problem and disregarding toroidality effects, the authors
derived the general kinetic equation for the distribution function of particles
$f(r,u,\tilde{ \mu })$ averaged in the ensemble of fluctuations, which, by
virtue of cylindrical symmetry about the angle $\theta$  and the direction $z$
along the cylinder axis  depends only on the radius  $r$ \cite{21}:
\begin{equation} \frac{\partial f}{\partial t} + \frac{e}{m_{e}}\, E_{e}
\frac{\partial f}{\partial u}= St(f) + I(f) \label{eq:3} \end{equation}

\noindent
where the collision integral of particles with fluctuations is given by
\begin{eqnarray}
&I(f)=\frac{u}{B(r)} \frac{1}{r} \frac{\partial}{\partial r} \{ rK \}
+\frac{\partial}{\partial u} \{ (\frac{e}{m_{e}} \frac{E_{a}}{B} -
\frac{\tilde{\mu}}{2}   \frac{dB}{dr}
\frac{1}{B} )K  \} & \nonumber \\[7pt]          \displaystyle
&K=\frac{|u|}{u}   \frac{F}{B} \frac{\partial f}{\partial r}+ \frac{F}{|u|}
\frac{\partial f}{\partial u} \{ \frac{e}{m_{e}}
\frac{E_{a}}{B} - \frac{\tilde{\mu}}{2}
\frac{dB}{dr}\frac{1}{B} \}
     \nonumber &
\end{eqnarray}

\noindent
Here the ambipolar electric field
$eE_{a}=(\frac{dn}{dr}\, \frac{T_e}{n} \, +\frac{1}{2}\frac{dT_e}{dr})
 \{ 1- 2\delta_{m} \}$
due to the difference between the
diffusion rates of electrons and ions, the constant
$\delta_{m}=\sqrt{m_{e}/m_{i}}\ll 1$ gives correction related to
ion diffusion
(assuming that transport of ions is  defined by magnetic fluctuations too),
$n(r)$  is the particle number
density,
$F=\int^{\infty}_{0} dL \langle b_{r} b_{r}^{'}
\rangle $
-is the correlation function of fluctuations $b_{r}$,  the
integration over  $L$  goes along the trajectory of particle motion,
$b_{r}^{'}=b_{r}(r^{'}(L),\theta^{'}(L),
z^{'}(L))$,
$u$  is the particle velocity along the magnetic field line,
$\tilde{\mu}=u_{\perp}^{2}/B$  is the
adiabatic invariant of particle motion,  $E_{e}$  is the external field,
$E_{e}={\vec {E_{e}}\cdot\vec{h}}, \, \,
 {\vec{h}}=\vec {B_{0}}/B_{0}$,
$St(f)$- is the Coulomb collision integral of particles.

In plasma
heating devices, the number of fast particles  $N_{f}$  is always small as
compared to the concentration  $N$  of main particles.  That is why the
kinetic equation (\ref{eq:3}) can be linearized in the small parameter
$N_{f}/N\ll 1$.
Equation  (\ref{eq:3}) is convenient to write in a spherical coordinate system
in the velocity space  $\epsilon , \, \mu$, where
$\epsilon =u^{2}+ u^{2}_{\perp
}$
is the total energy and
$\mu =u/\sqrt{u^{2}+ u^{2}_{\perp }}$ is the cosine of
pitch angle (below for brevity we'll use simply pitch-angle without cosine).
Linearizing we obtain
\begin{eqnarray} \label{eq:4} &\frac{\partial f}{\partial
\tau} +
E_{e}(r)\delta_{1} \{ 2\mu \epsilon \frac{\partial f}{\partial \epsilon}
+ (1-\mu^{2}) \frac{\partial f}{\partial \mu} \} = 4T(r)\frac{\partial^{2}
f}{\partial \epsilon^{2}}
+ 2\frac{\partial f}{\partial \epsilon}+ \nonumber \\
&\frac{Z_{eff}}{\epsilon} \, \frac{\partial }{\partial \mu} \{
(1-\mu^{2})\frac{\partial f}{\partial \mu } \} +\epsilon \mu \delta_{2} \{
\frac{1}{rB} \frac{\partial}{\partial r} \{ rK \}  +   \nonumber \\
&\frac{2E_{a}}{B} \{  \frac{\partial K}{\partial \epsilon} +
\frac{1-\mu^{2}}{2\epsilon\mu} \frac{\partial K}{\partial \mu} \} -
\frac{dB}{dr}     \frac{1}{B^{2}}
\frac{1-\mu^{2}}{2\mu}
\frac{\partial K}{\partial \mu}     \}  \\
&K= \frac{|\mu|}{\mu}  \frac{F}{B}    \{
\frac{\partial f}{\partial r} +  2E_{a} \{
\frac{\partial f}{\partial\epsilon}+
\frac{1-\mu^{2}}{2\epsilon\mu}\frac{\partial f}{\partial\mu} \} -
\frac{dB}{dr}\frac{1}{B}
\frac{1-\mu^{2}}{2\mu}
\frac{\partial f}{\partial\mu} \} \nonumber
\end{eqnarray}

\noindent
Here  $E_{e}(r)$  is the profile of the external longitudinal electric field
normalized to the critical electron runaway field  $E_{c}$  \cite{gr1},
$E_{c}=\frac{4\pi e^{3}\Lambda n}{T_{e}}$, $\: Z_{eff}$
is the effective ion charge, $\Lambda$ is Coulomb logarithm,
$\delta_{1}=E_{e0}/E_{c}$ is the dimensionless parameter characterizing the
magnitude of the longitudinal field  $E_{e0}$ relative to the critical field
$E_{c}$, $\delta_{2}=\nu_{a}/\nu_{0}$  is a dimensionless
parameter characterizing the particle-fluctuation collision frequency
$\nu_{a}= \frac{F_{max}}{a^{2}B_{0}^2}\sqrt{ \frac{T_{e}}{m_{e}} }$
as compared with the Coulomb collision frequency of electrons
$ \nu_{0}=\frac{4\pi e^{4}n\Lambda}{m_{e}^{1/2}T_{e}^{3/2}}$, $a$
  is the characteristic system dimension, $T(r)$
- is the profile of the electron temperature normalized to the temperature at
the center $T_e$.
Furthermore, dimensionless quantities $r=\tilde{r}/a, \,
 F=\tilde{F}/F_{max}, \, B=\tilde{B}/B_{0}, \, E_{a}=\tilde{E_{a}}ea/T_{e}, \,
\epsilon =\tilde{\epsilon}/ (T_{e}/m_{e}), \, \tau =\nu_{0}t  $ are introduced,
the sign  \~{} marks the corresponding dimensional quantities, $\delta_{1}$
and
$\delta_{2}$ are small parameters of our problem.  In what follows we shall
consider steady-state solutions of equation (\ref{eq:4}) to determine the
established distribution function of superthermal particles.  We shall assume
the distribution function of the main particles in plasma to be stationary and
equilibrium.  In such a statement, it will be a source of superthermal plasma
particles.  As boundary conditions for the distribution function  $f$  it is
natural to require that  $f$  be regular as $r\rightarrow 0$   and that all the
particles die on the boundary for  $r = 1$, that is, $$\frac{\partial
f}{\partial r}|_{r=0}=0 \, \, \, f|_{r=1}=0$$

The investigation of the complete problem is difficult because of a
simultaneous action of such factors as an applied electric field  $E_{e}$, an
anomalous diffusion and  inhomogeneity of the mean magnetic field.  We
shall therefore begin with examining a joint effect of the external
electric field  $E_{e}$ and the fluctuations assuming the magnetic field
gradient to be small

\begin{equation}
\frac{dB}{ dr}\frac{a}{B}\ll 1
\label{eq:6}
\end{equation}

\noindent
and then proceed to the case of an inhomogeneous magnetic field  $B(r)$.


\section{ Distribution function of superthermal
    electrons in a homogeneous magnetic field.}

Provided that the condition  (\ref{eq:6})  is satisfied,  equation
(\ref{eq:4})
has the form
\begin{eqnarray}
\label{eq:7}
& E_{e}(r)\delta_{1} \{
2\mu \epsilon \frac{\partial f}{\partial \epsilon}  + (1-\mu^{2})
\frac{\partial f}{\partial \mu} \} =
4T(r)\frac{\partial^{2} f}{\partial \epsilon^{2}}  +
2\frac{\partial f}{\partial
\epsilon}+ \nonumber \\
&\frac{Z_{eff}}{\epsilon} \, \frac{\partial }{\partial \mu} \{
(1-\mu^{2})\frac{\partial f}{\partial \mu } \}
+\epsilon \mu \delta_{2} \{ \frac{1}{rB}   \frac{\partial}{\partial r}
\{ rK \}  +   \nonumber \\
&2E_{a} \{  \frac{\partial K}{\partial \epsilon} +
\frac{1-\mu^{2}}{2\epsilon\mu} \frac{\partial K}{\partial \mu} \}
     \}  \\
&K= \frac{|\mu|}{\mu}  F    \{
\frac{\partial f}{\partial r} +  2E_{a} \{
\frac{\partial f}{\partial\epsilon}+
\frac{1-\mu^{2}}{2\epsilon\mu}\frac{\partial f}{\partial\mu} \}
 \} \nonumber  \\
&E_{a}=(\frac{dn}{dr}\frac{T(r)}{n}+\frac{1}{2}\frac{dT}{dr})
\{ 1-\delta_{m}   \} , \, \, \delta_{m}\ll 1 \nonumber
\end{eqnarray}

\noindent
Equation  (\ref{eq:7})  describes the established distribution function of
electrons in the presence of the field  $E_{e}$  and plasma turbulence.
The method of
solving equation  (\ref{eq:7})  depends on the energy range within which we
seek
the solution.  Therefore we shall first examine the energy range immediately
adjoining the equilibrium region, which is henceforth referred to as a
polynomial region of solution

\begin{equation} \label{eq:8}
1 \leq \epsilon \leq \delta^{-1/2}_{1} , \,\, \delta_{1}\ll 1
\end{equation}

We shall not
consider here the thermal runaway of particles  \cite{26,26add}.
According to the results of \cite{26add} one may estimates the
critical value of energy $y_k$, $y=\epsilon \delta_1^{1/2}$
when the temperature profile
will relax to homogeneous one,
$y_k=\delta_1^{1/2}\delta_2^{-1/3}
(\frac{a}{L_{\| c}})^{2/3}$, where $a$ - is a scale of  the system,
$L_{\| c}$ - is a
correlation length of fluctuations along magnetic field line.
As usually $\frac{a}{L_{\| c}}\ll 1$ and we assume that,
\begin{equation}
\label{cond}
y_k\ll 1
\end{equation}

And therefore, in accordance with (\ref{cond})
in equation(\ref{eq:7}) we have put
$T_{e}=const$ and
for the sake of simplicity $T_{i}=const$, because ion temperature
profile contributes only to a correction term in ambipolar field.
So, as is seen inhomogeneity of electron temperature contributes only
to ambipolar field.

\subsection{ The behavior of solution in the polynomial region}

To begin with, we consider the case of a constant field  $E_e(r) = const$.
We imply at first to clear up the effects of ambipolar field,
that $\frac{dT}{dr}\frac{a}{T}\ll \delta_m$ and neglect contribution
from temperature gradient in ambipolar field.
In equation (\ref{eq:7})  we pass over to a new variable
$y=\epsilon\delta^{1/2}_{1}$  and seek the solution
as a series of eigenfunctions of the Sturm-Liouville problem
\begin{eqnarray}
\label{eq:9}
&\frac{1}{r}\frac{\partial }{\partial r} \{rF
\frac{\partial\chi_{l}}{\partial r} \}+\eta_{l}\chi_{l} =0  \\
&\frac{\partial\chi_{l}}{\partial r}|_{r=0}=0        \nonumber     \\
&\frac{\partial\chi_{l}}{\partial r}|_{r=1}=0        \nonumber
\end{eqnarray}
that is,
\begin{equation}
\label{eq:10}
f=C\, n(r)\exp (-\Psi_{0}/\delta_{1}^{1/2})\sum_{l}\chi_{l}(r)R_{l}(y,\mu )
\end{equation}
As the boundary condition for  $y\rightarrow 0$  we require the condition of
sewing with
the equilibrium distribution function
$$
f_{0}=C\, n(r)\exp (-y/\delta_{1}^{1/2})$$.
$$n(r)|_{r=1}=0$$

\noindent
Substituting  (\ref{eq:10})  into  (\ref{eq:7})  and collecting terms with the
same power  $\delta^{1/2}_{1} $  in zero approximation, we obtain
$\Psi_{0}=y/2$.  In the next order of
perturbation theory, neglecting correction of the order of
$\delta_{m}\ll 1$ in the expression
for the ambipolar field    (\ref{eq:7}),  we obtain a system of equations for
the functions  $R_{l}(y,\mu )$
\begin{equation}
\label{eq:11}
2\frac{\partial R_{l}}{\partial y}=\frac{Z_{eff}}{y}\frac{\partial
} {\partial\mu} \{ (1-\mu^{2})
\frac{\partial R_{l}}{\partial\mu} \} -y|\mu |\beta\eta_{l}R_{l}+ 2\mu y R_{l}
\end{equation}
with boundary conditions as $y\rightarrow 0$
$$R_{0}(0,\mu )=1, \, \, R_{l}(0,\mu )=0 \: \: l\neq 0 $$,

\noindent
where
$ \beta
=\delta_{2}/\delta_{1}$.
Since zero eigenvalue of the problem  (\ref{eq:9}) $\eta_{0}=0$,
 for the function  $R_{0}(y,\mu )$
we obtain an equation containing no contribution of anomalous diffusion,
and it will have a solution similar to the one obtained in the paper
\cite{gr1}, where the distribution function was distorted only by the electric
field  $E_e$  and had a directed character for  $\mu\simeq 1$  (Fig. 1, the
dashed line). The solution of the system  (\ref{eq:11}) for  $l\neq 0$, which
satisfies the boundary conditions  (because for  $l\neq 0$  all the
eigenvalues  $\eta_{l}>0$  are nonnegative),  will be  $R_{l}(y,\mu )=0$.
So,  up to terms of the
order of  $\delta^{1/2}_{1}$  (terms of the order of
$\delta^{1/2}_{1}\delta_{m} \ll 1$  and
$\delta_{1}\ll 1$  are neglected)  the
initial equilibrium distribution function  $f_{0}$  will not be distorted by
anomalous
transport.  This means, as is readily seen, that the contribution from
the anomalous diffusion is completely compensated by the ambipolar field
$E_a$  in exactly the same way as in the case of diffusion of thermal plasma
particles.  As is well known, electrons and ions diffuse together as a
single whole with a doubled ion diffusion coefficient, which in our case
is  $\delta_{m}$  times smaller than the electron one  (provided that the
predominant ion diffusion mechanism is also an anomalous transport
caused by magnetic fluctuations).

To determine the distortion of the
distribution function by an anomalous transport,  we have to examine the
solution of equation  (\ref{eq:7})  with allowance for terms of the order of
$\delta^{1/2}_{1}\delta_{m}$
and  $\delta_{1}$
,  which under certain conditions  (to be discussed below)  become
appreciable.  Now we are in a position to consider the case where the
parameter  $\delta_{m}$  satisfies the condition
$\delta_{m}\gg\delta^{1/2}_{1}$.  This means that we shall take
into account the contribution from the anomalous diffusion, proportional
to the anomalous ion diffusion coefficient.

With allowance for
corrections of the order of  $\delta_{m}$, the solution of equation
(\ref{eq:7})
is convenient to seek as before in the form of an eigenfunction series of the
Sturm-Liouville problem
\begin{eqnarray}
& \label{eq:12}
\frac{1}{r}\frac{\partial }{\partial r} \{rF
\frac{\partial g_{l}}{\partial r} \}+\lambda_{l}g_{l} =0  \\
&\frac{\partial g_{l}}{\partial r}|_{r=0}=0        \nonumber     \\
& g_{l}|_{r=1}=0        \nonumber  \\
&\label{eq:13}
f=C\, \exp (-\Psi_{0}/\delta_{1}^{1/2})\sum_{l}g_{l}(r)\Theta_{l}(y,\mu ) \\
&\nonumber n(r)=\sum_{l}A_{l}g_{l}
\end{eqnarray}

\noindent
with the boundary condition as  $y\rightarrow 0$:
\begin{equation}
\label{eq:14}
\Theta_{l}(0,\mu ) =A_{l}
\end{equation}

In order that we might pass over to a system of ordinary differential
equations,
it is also convenient to expand the functions
$\Theta_{l}(y,\mu )$  in a power series of Legendre
polynomials  $P_{m}(\mu )$:
\begin{equation}
\label{eq:15}
\Theta_{l}(y,\mu ) =\sum_{m}F_{l}^{m}(y)P_{m}(\mu )
\end{equation}

Substituting (\ref{eq:13}) into (\ref{eq:7}) with account of (\ref{eq:15}), we
ultimately arrive at a system of linking  equations for the functions
$ F_{l}^{m}(y)$,
$\Psi_{0}=y/2$:
\begin{eqnarray} & \label{eq:16}
2\frac{dF^{m}_{l}}{dy}= -\frac{Z_{eff}}{y}m(m+1)F^{m}_{l}
+yF_{k}^{n}\alpha_{kl}\gamma_{mn}\beta +\\ &2y\{
\frac{F^{m-1}_{l}}{2m-1}+\frac{(m+1)F^{m+1}_{l}}{2m+3} \} \nonumber
\end{eqnarray}
with boundary conditions
\begin{eqnarray}
& F^{0}_{l}(0)=A_{l} \nonumber \\ & F^{m}_{l}(0)=0 \: \: m\neq 0
\nonumber \end{eqnarray}

\noindent
where
\begin{eqnarray}
& \nonumber \alpha_{kl}=\int^{1}_{0}rg_{l}\{
\frac{1}{r}  \frac{\partial }{\partial r } \{ r\Pi_{k} \} -E_{a}\Pi_{k} \} dr
\,
/\int^{1}_{0}rg^{2}_{l}dr \\ &\Pi_{k}=F\{ \frac{\partial g_{k}}{\partial r}
-E_{a}g_{k}  \} \nonumber \\ & \gamma_{mn}=(2m+1)\int^{1}_{-1}P_{m}P_{n}|\mu
|d\mu , \: \: \beta = \delta_{1}/\delta_{2}    \nonumber
\end{eqnarray}

The system of equations  (\ref{eq:16})  is solved numerically by cutting off
the
chain of equations on the term  $L$  in the expansion in  $g_{l}$  and
on the term  $M$
in the expansion in  $P_{m}$, so that a doubling of  $L$  and  $M$   changes
the
solution by less than 10 \%.  Clearly, strong distribution function distortions
by an anomalous transport will take place only when the parameter
$\Delta =2\beta\lambda_{m}\delta_{m} $  is of the
order of unity  $\Delta\sim 1$,  where
$\lambda_{m} $  is the maximum eigenvalue corresponding to the
eigenfunctions $g_{l}(r)$  which contribute to the expansion  (\ref{eq:14}).
The smallest distortions may be expected in the case $n(r)=g_{0}(r)$.

We are examining the
behavior of the system (\ref{eq:16}) in the model case $F = const$, when the
eigenfunctions of the problem (\ref{eq:12}) are the Bessel functions
$g_{l}(r)=J_{0}(\xi_{l}r)$.  We
begin with examining the case  $n(r)=g_{0}(r)$.
The solution of the system (\ref{eq:16}) depending on
the pitch angle  $\mu$  and the energy  $y$  is presented in Figs. 1, 2.
Figure
1 shows solutions obtained for various values of the parameter $\beta$.
For small $\beta$,
when  $\Delta\ll1$  and the effect of the electric field  $E_e$  dominates,
the
solution is close to that obtained in the paper \cite{gr1} (the dashed line in
Fig. 1).  As  $\beta$ increases, the influence of the anomalous transport
(when
$\Delta\geq 1$) becomes predominant.
In this case the solution becomes almost symmetric
in  $\mu$.  As should be expected,  it is concentrated in the vicinity of
$\mu =0$
, where the diffusion coefficient  (\ref{eq:2})  vanishes and falls
symmetrically for  $\mu\Rightarrow \pm 1$,
where the diffusion coefficient is maximal.
Depending on the energy  (Fig.2), the distribution function falls exponentially
 $\ln(f/f_{0})\sim-\Delta_{0}y^2$,
where  $ \Delta_{0}=2\delta_{m}\beta\lambda_{0}$
($\lambda_{0}$  is a nonzero eigenvalue of the problem (\ref{eq:12})).
The function $f$, which is shown in fig.2, was averaged over $\mu$.

It should be emphasized that the initial
distribution over discharge  $n(r)$ will not be deformed considerably.

We shall now see what will happen if as the initial
profile   $n(r)$  we shall choose an arbitrary function $\tilde{n}(r)$
such that the expansion
(\ref{eq:14}) will involve (and with a substantial contribution) harmonics with
$l\neq 0$, as is shown in Fig.3.  The harmonics with eigenvalues
$\lambda_{l}\gg \lambda_{0}$   should be
expected to damp faster than the zero component does already for  $y<1$.  As a
result,  the distribution over discharge will rather rapidly relax to
$g_{0}(r)$, which
will immediately lead to balance violation between the ambipolar field and
anomalous diffusion.  In this case, an effective increase of  the diffusion
rate
may be expected.  Indeed, Fig.3 shows the distribution over discharge radius
for
$y = 1$,
obtained in the solution of the system (\ref{eq:16}) with the profile  $n(r)=
\tilde{n}(r)$,
is almost coincident with  $g_{0}(r)$ (dashed line in the same figure).
On the
other hand, Fig.2 presents the dependencies obtained for one and the same value
of $\beta$ for  $n(r) = g(r)$  and the profile  $n(r)=\tilde{n}(r)$  depicted
in
Fig. 3.  It is seen that in the latter case the decrement is considerably
larger, which testifies to an effective increase of the diffusion rate.

The result was obtained make allowance us to conclude, that the rate and
character of diffusion (and as a consequence lost of energy due to diffusion of
fast particles) essentially depends on  profile of mean particle density.
The minimum of lost will be as   $n(r)=g_{0}(r)$.

The density profile in a device $n(r)$ is defined by many factors:
anomalous transport, neutral particles flow from the camera wall and external
sources, accelerated ions injection, convective transport processes.
So by means of injection, for example, one may to drive profile $n(r)$ and
respectively process of fast particle transport.

The result obtained
suggests that the compensation of anomalous transport by an ambipolar field is
a
consequence of equilibrium in the leading term of the distribution function of
electrons, that is  $\Psi_{0}(y)=y/2$,
and the absence of distortion of the initial electron
distribution $n(r)$.  As  shown above,  the strict balance is violated by the
anomalous transport itself when $n(r)\neq g_0(r)$.  In the presence
of an electric field such a balance will also be violated for  $y >> 1$
 since according to  \cite{gr1} the electric field  $E_e$
induces strong deviations of electron distribution from an equilibrium one.  We
may point out another mechanism of balance violation already for  $0 < y < 1$,
namely, inhomogeneity of the applied field   $E_{e}(r)$  which in this case
plays the role of a fast particle source nonuniform in space  (note that it is
exactly the case realized in  RFP).  We shall consider the solution of equation
(\ref{eq:7}) in the polynomial region using the methods presented above.
As before, we put  $F = const$.  Figure 2 presents the dependence of the
distribution function $\ln (f/f_{0})$
on the energy  $y$  for one and the same value of the parameter  $\beta$ and
$n(r) = g_0(r)$  in two cases:

\noindent
{\it a)} for  $E  = const$  and  {\it b)} for
$E_{e}(r)=J_{0}(\kappa_{0}r),\:\kappa_{0}=3.0$,
(such a profile is close to the one observed in
RFP discharges).  It can be readily seen that in case  {\it b)}
the
distribution function fall with energy is much higher than in case {\it a)},
which shows an acceleration of the diffusion process.
It is noteworthy that the
electron temperature inhomogeneity (which we do not consider here) will
obviously play a role similar to that played by the inhomogeneity of the
external field  $E_e(r)$  and will also lead to a violation of strict balance,
see Fig.2.

We have assumed above that  $\delta_{m}\gg\delta_{1}^{1/2}$,
and the correction of the order of  $\delta_{m}\delta_{1}^{1/2}$ has led
  to strong distribution function distortion when
$\Delta\geq 1$.
Clearly,  in the converse case, that is when
$\delta_{m}\ll\delta_{1}^{1/2}$, there will proceed an
analogous process which is not distinct qualitatively from the one investigated
above with the only difference that in this case the role of the parameter
$\Delta$
will be played by the parameter
$\Delta_{1}=2\beta\lambda_{m}\delta_{1}^{1/2}$ .
As a result, as before for  $\Delta_{1}\sim 1$  the
distribution function will relax rapidly to the profile  $g_0(r)$.

Thus we have completely investigated  the
behavior of the solution of equation (\ref{eq:7}) in the polynomial region.
We
have obtained that in a special case where the profile  $n(r)$  is chosen in
the
form $n(r) = g_0(r)$  and the external field  $E_e$   is homogeneous, the
ambipolar field $E_a$  completely damps the anomalous transport of fast
electrons with an accuracy of $\delta_{m}\ll 1$, so that the effective
diffusion is determined by the parameter $\Delta$ or $\Delta_1$ which is
substantially smaller than $\beta\lambda_0, \:\Delta\ll\beta\lambda_0$.
On the other hand, such a strict balance in
higher terms of expansion may be violated provided that $n(r)\neq g_0(r)$ or
the external field $E_e$ and temperature $T(r)$ are inhomogeneous.
We note that the latter always takes place in  RFP.  An
essential result is here the fact that for  $\Delta >1$  (or  $\Delta_{1}>1$)
the distribution function of fast particles over discharge relaxes with
increasing energy to the universal profile $g_0(r)$ independent of the initial
distribution.  This fact will simplify appreciably our analysis in the
remaining
part of this section, where we consider the energy range  $y >> 1$.

\subsection {The behavior of the solution in the exponential region.}

We shall now consider the domain of solution for high energy values
$\epsilon \gg \delta^{-1/2}$, where
$\delta=\delta_{1},\delta_{2}$.
The distortions of the distribution function in this domain are known
to be of exponential character \cite{23a}.
Therefore, a polynomial expansion is not
effective here.  As we have seen above, practically for
$\epsilon \gg\delta^{-1/2}$ , the distribution
function of fast electrons over discharge is coincident with the zero
eigenfunction  $g_0(r)$ of the problem  (\ref{eq:12}).  Therefore, in this
energy range, in the expansion of the distribution function it is natural to
make allowance only for terms containing  $g_0(r)$:
\begin{equation}
\label{eq:17}
f=C\sum_{l}g_{l}\exp{(-\Psi_{l})}
\end{equation}

Let us consider the case  $E_{e}=0$.  We shall pass over to a new variable
$z=\epsilon\delta _{2}$
and represent  $\Psi$  as:
\begin{equation}
\label{eq:17.1}
\Psi=\Psi_{0}/\delta_{2}+\Psi_{1}/\delta_{2}^{1/2}+\Psi_{2}+\ldots
\end{equation}

\noindent
Substituting  (\ref{eq:17}) and (\ref{eq:17.1}) into (\ref{eq:7}) and
collecting
terms with the same powers  $\delta_{2}^{1/2}$,
we obtain the system of equations
\begin{equation}
\label{eq:18.1}
\frac{\partial\Psi_{0}}{\partial\mu}=0
\end{equation}

\begin{equation}
\frac{\partial\Psi_{0}}{\partial\mu}\frac
{\partial\Psi_{1}}{\partial\mu}=0 \label{eq:18.2}
\end{equation}

\begin{equation}
4(\frac{\partial\Psi_{0}}{\partial z})^{2} -
2\frac{\partial\Psi_{0}}{\partial z}+ \frac
{Z_{eff}}{z}(1-\mu^{2})(\frac{\partial\Psi_{1}}{\partial \mu})^{2} -
z|\mu |\{ \lambda_{0}+B(\frac{\partial\Psi_{0}}{\partial z})^{2} \} = 0
 \label{eq:18.3}
\end{equation}      $$B=-4\int_{0}^{1}
rg_{0}^{2}(r)F(r)E_{a}^{2}(r)dr /\int_{0}^{1} rg_{0}^{2}(r)dr $$

\noindent
{}From  (\ref{eq:18.1})  it follows that $\Psi_{0}=\Psi_{0}(z)$,
and (\ref{eq:18.2})  holds identically.  From the
condition of the absence of a jump of the derivative
$\frac{\partial\Psi}{\partial\mu} $  for $\mu =0$
there follows a natural condition  (in view of symmetry under a
substitution of  $\mu$  for  $-\mu$  in  (\ref{eq:7})  for  $E_e= 0$)
$$
\frac{\partial\Psi_{1}}{\partial\mu}|_{\mu=0}=0$$
Taking into consideration this condition, as well as the fact that
$\Psi_{0}=\Psi_{0}(z)$, from
equation  (\ref{eq:18.3})  for $\mu =0$  we obtain
$$\Psi_{0}(z)=z/2$$
Substituting the expression found for $\Psi_{0}(z)$   back into
(\ref{eq:18.3}),
we obtain the equation for determining  $\Psi_{1}(z,\mu )$.
\begin{equation}
\frac{\partial\Psi_{1}}{\partial\mu}=z\sqrt{\frac{|\mu|\lambda_{0}^{*}}
{z_{eff}(1-\mu^{2})}}
\label{eq:19}
\end{equation}

\noindent
where  $\lambda_{0}^{*}=\lambda_{0}+B/4$  is an effective
eigenvalue with account of the influence of the
ambipolar field  $E_a$.  From (\ref{eq:19}) we see however that the
asymptotical
expansion obtained is violated in the vicinity of  $\mu = 0$ because the second
derivative  $\frac{\partial^{2}\Psi}{\partial\mu^{2}} $
contains a singularity.  Indeed,
$\frac{\partial^{2}\Psi}{\partial\mu^{2}} \rightarrow\infty$
$\mu\rightarrow 0 $,
which indicates of the
presence of a boundary layer near  $\mu = 0$.  To obtain a correct expansion,
it
is necessary to investigate the behavior of the solution in the vicinity of
$\mu =0$.
To this end, the small term with a second derivative
$\frac{\partial^{2}\Psi}{\partial\mu^{2}} \delta_{2}^{1/2} $ in equation
(\ref{eq:18.3}) should be retained.  Omitting terms of the order of
$\mu^{2}$  as  $ \mu\rightarrow 0$
in (\ref{eq:18.3})  and making a substitution
$\Psi_{1}=-\ln (\Theta )\delta_{2}^{1/2}$, we come to the Airy equation
\begin{equation}
\frac{\partial^{2}\Theta}{\partial\xi^{2}}=\xi\Theta
\label{eq:20}
\end{equation}
$$\xi=\frac{\lambda^{*}z^{2}\mu-\varphi(z)}{z_{eff}^{1/3}z^{4/3}
\delta_{2}^{1/3} {\lambda^{*}}^{2/3} }$$
$$\varphi
(z)=\{4(\frac{\partial\Psi_{0}}{\partial z})^{2} -
2\frac{\partial\Psi_{0}}{\partial z}\}
$$
$$\lambda^{*}=\lambda_{0}+B(\frac{\partial\Psi_{0}}{\partial z})^{2}$$

\noindent
Equation  (\ref{eq:20})  has two linearly independent fundamental solutions,
one
of which grows exponentially as  $\xi\rightarrow\infty$ and
may be discarded for being limited.  The
other solution  $\Theta=C_{0}Ai(\xi)$  is shown in Fig. 4,
$\xi_{0}$  is the point where the function  $\Theta$
and  $\Psi_{1}$ has an extremum,
$ \frac{\partial\Psi_{1}}{\partial\mu}=0$.
Then from the condition
$\frac{\partial\Psi_{1}}{\partial\mu}|_{\mu=0}=0$  we obtain an additional
relation that allows us to determine the function  $\Psi_{0}(z)$
\begin{equation}
4(\frac{\partial\Psi_{0}}{\partial z})^{2} -
2\frac{\partial\Psi_{0}}{\partial
z}=|\xi_{0}|z_{eff}^{1/3}{\lambda^{*}}^{2/3}(z)
\delta_{2}^{1/3}z^{1/3}
\label{eq:21}
\end{equation}

In the general case, to find  $\Psi_{0}(z)$
it is necessary to solve the nonlinear equation
(\ref{eq:21}), but for  $z\ll\delta^{-1}_{2}$
expanding equation (\ref{eq:21}) in the small parameter  $\delta_{2} $ we
obtain
\begin{equation}
\Psi_{0}(z)=\frac{z}{2}+\frac{3}{4}|\xi_{0}|z_{eff}^{1/3}{\lambda_{0}^{*}}^{2/3}\delta_{2}^{1/3}
z^{4/3}
\label{eq:22}
\end{equation}
$$\lambda_{0}^{*}=\lambda_{0}+B/4$$
Taking into account (\ref{eq:22}), we may determine the angular dependence of
the distribution function:
\begin{equation}
\Psi_{1}(\mu,z)=\int^{\mu}_{\mu_{0}}
\sqrt{\frac
{z\{ \lambda^{*}(z)z|\mu|-|\xi_{0}|z_{eff}^{1/3}{\lambda^{*}(z)}^{2/3}z^{1/3}
\delta^{1/3}_{2}\}}
{z_{eff}(1-\mu^{2}) } } d\mu    + \tilde{\Psi}_{1}(z)
\label{eq:23}
\end{equation}
where  $\tilde{\Psi}_{1}(z)$  is an unknown function.
As  $\mu\rightarrow 0$, it is necessary to sew (\ref{eq:23}) with
the solution of equation (\ref{eq:20}) to find the constant  $C_0$  .  It is
readily seen that as  $\delta_{2}\rightarrow0$,
(\ref{eq:22}) transforms into  $\Psi_{0}(z)=z/2$  and (\ref{eq:23})  into
(\ref{eq:19}).  As  $z\rightarrow 0$, the solution found here must pass over to
the solution
obtained in the polynomial region.  Figure 5 shows dependencies of the
distribution function on the pitch angle in the polynomial region for  $y > 1$
and the solution obtained in the exponential region when  $z\rightarrow 0$.
One can see
good agreement between the two solutions.  The solution obtained by us shows
that the leading term of the asymptotical expansion for
$\delta_{2}\rightarrow 0$  is close to
equilibrium, and deviations from it occur only for
$z\sim\delta_{2}^{-1}$.  This result is in close
agreement with the results of the direct numerical simulations
\cite{28}  carried out
for  TOKAMAK  conditions in the absence of the applied electric field  $E_e$
 and experiments \cite{28a},\cite{28aa}.
It should be noted that the role of an ambipolar field in this range of energy
values comes down to renormalization of the eigenvalue
$\lambda_{0}$  which determines the
anomalous transport, namely, the effective eigenvalue with allowance for the
influence of the ambipolar field  $E_a$  will be
$\lambda^{*}_{0}=\lambda_{0}+B/4$ , $\lambda^{*}_{0}<\lambda_{0}$
since  $B < 0$.  That is, as
expected, the ambipolar field  $E_a$  damps the anomalous transport of fast
electrons. It is clear that, this effect is important when the effective
temperature of fast electrons is of the order of equlibrium temperature at the
center of the discharge.

The field  $E_a$  also affects the behavior of the distribution
function in the far region of energy values as  $z\rightarrow \infty$.
So, when  $B\neq 0$  one
can readily obtain an asymptotical expression for $\Psi_{0}(z)$  as
$z\rightarrow\infty$:
$$
\Psi_{0}(z)\simeq z\sqrt{\lambda_{0}/B}
$$

At the same time when  $B = 0$ and  $z\rightarrow\infty$, from (\ref{eq:21}) we
have
$$
\Psi_{0}(z)\simeq
\frac{3}{14}z^{7/6}\lambda_{0}^{1/3}z_{eff}^{1/6}\delta_{2}^{1/6} $$

\noindent
that is, in the absence of the ambipolar field  $E_a$  the distribution
function
falls stronger with energy.

We shall now consider the case with a
nonzero external field  $E_e$.  As in the previous case in the expansion
(\ref{eq:17}),  we shall take into account only the contribution from the terms
$g_{0}(r)$.  In equation (\ref{eq:7}) we pass over to a new variable
$z=\epsilon\delta_{1}$ and
represent the index of the exponential in (\ref{eq:17}) in the form
\begin{equation}
\Psi=\Psi_{0}/\delta_{1}+\Psi_{1}/\delta_{1}^{1/2}+\Psi_{2} \label{eq:26}
+\ldots
\end{equation}
Substituting (\ref{eq:26}) into (\ref{eq:7}) and collecting terms with the same
powers  $\delta_{1}^{1/2}$, we obtain a chain of connected equations for the
functions
$\Psi_{0},\Psi_{1},\Psi_{2},\ldots$ :
\begin{equation}
\frac{\partial\Psi_{0}}{\partial\mu}=0
\label{eq:27.1}
\end{equation}

\begin{equation}
\frac{\partial\Psi_{0}}{\partial\mu}\frac
{\partial\Psi_{1}}{\partial\mu}=0
\label{eq:27.2}
\end{equation}

\begin{equation}
4(\frac{\partial\Psi_{0}}{\partial z})^{2} -
2\frac{\partial\Psi_{0}}{\partial z}+ \frac
{Z_{eff}}{z}(1-\mu^{2})(\frac{\partial\Psi_{1}}{\partial \mu})^{2} -
z|\mu |\beta\lambda^{*}(z)+2\mu zE\frac{\partial\Psi_{0}}{\partial z} = 0
 \label{eq:27.3}
\end{equation}

\begin{eqnarray}
8\frac{\partial\Psi_{0}}{\partial z}\frac{\partial\Psi_{1}}{\partial z}  -
2\frac{\partial\Psi_{1}}{\partial z} +\frac{Z_{eff}}{z}
\{2(1-\mu^{2})  \frac{\partial\Psi_{1}}{\partial
\mu}\frac{\partial\Psi_{2}}{\partial
\mu}-\frac{{\partial}^{2}\Psi_{1}}{\partial
\mu^{2}}  +2\mu\frac{\partial\Psi_{1}}{\partial \mu}  \}  -
&\nonumber \\ 2z|\mu|\beta B\frac{\partial\Psi_{0}}{\partial
z}\frac{\partial\Psi_{1}}{\partial z} +2\mu zE\frac{\partial\Psi_{1}}{\partial
z} + E(1-\mu^{2})\frac{\partial\Psi_{1}}{\partial \mu }  =  0 &
\label{eq:27.4}
\end{eqnarray}
$$B=-4\int_{0}^{1} rg_{0}^{2}(r)F(r)E_{a}^{2}(r)dr /\int_{0}^{1}
rg_{0}^{2}(r)dr $$
$$E=\int_{0}^{1} rg_{0}^{2}(r)E_{e}(r)dr /\int_{0}^{1} rg_{0}^{2}(r)dr $$
$$ \beta =\delta_{2} / \delta_{1} ; \; \; \lambda^{*}(z) =\lambda_{0}+
B(\frac{\partial\Psi_{0}}{\partial z})^{2}$$

{}From the first equation  (\ref{eq:27.1}) it follows that
$\Psi_{0}=\Psi_{0}(z)$.  Given this,
equation (\ref{eq:27.2}) holds automatically.  Then, in the absence of a
singularity for  $\mu =1$, from (\ref{eq:27.3}) we obtain the equation with the
help of which we may determine $\Psi_{0}$:
\begin{equation} 4(\frac{\partial\Psi_{0}}{\partial z})^{2} -
2\frac{\partial\Psi_{0}}{\partial z} -z\beta\lambda^{*}(z)+2
zE\frac{\partial\Psi_{0}}{\partial z} = 0 \label{eq:28}
\end{equation}
This implies
\begin{equation} \frac{\partial\Psi_{0}}{\partial z}
=\frac{(1-zE)+\sqrt{(1-zE)^{2}+z\beta\lambda_{0} (4-\beta zB)}} {4-z\beta B}
\label{eq:29} \end{equation}
Integrating (\ref{eq:29})
we arrive at
\begin{eqnarray}
&\Psi_{0}(z)=\frac{Ez}{B\beta}-\frac{\sqrt{1+a_{1}z+a_{2}z^{2}}}{B\beta} -
\nonumber &\\
&-\frac{(8a_{2} + a_{1} B\beta) \ln( a_{1} + 2a_{2}z +
2\sqrt{a_{2}}\sqrt{1+a_{1}z+a_{2}z^{2} } ) } { 2\sqrt{a_{2}} B^{2}\beta^{2} } +
 \label{eq:30} &\\
&\frac{(4E-B\beta)\ln (S_{1} )}{B^{2}\beta^{2}} \nonumber &
\end{eqnarray}

$$
S_{1}=-4a_{1}B^{3}\beta^{3}-2B^{4}\beta^{4}-8a_{2}B^{3}\beta^{3}
z-a_{1}B^{4}\beta^{4}z-2B^{3}\beta^{3}(4E-B\beta)\times
$$

$$
\times\sqrt{1+a_{1}z+a_{2}z^{2} }
$$

$$
a_{1}=4\beta\lambda_{0}-2E  ; \; \; a_{2}=E^{2}-\beta^{2}\lambda_{0} B
$$
Taking into consideration (\ref{eq:29}), we
obtain from (\ref{eq:27.3})
\begin{equation}
\Psi_{1}(\mu,z)=-z\int^{\mu}_{1}
\sqrt{\frac
{
\beta
(\lambda_{0}+B(\frac{\partial\Psi_{0}}{\partial z})^{2})(|\mu|-1)-2
\frac{\partial\Psi_{0}}{\partial z}E(\mu -1)
} {z_{eff}(1-\mu^{2}) } } d\mu    + \tilde{\Psi}_{1}(z)
\label{eq:31}
\end{equation}
It should be noted that  (\ref{eq:31})  holds for
\begin{equation}
\beta\{ \lambda_{0}+B/4 \} < E
\label{eq:31.1}
\end{equation}
In this case, the solution will not have a singularity for  $\mu =1$.
Substituting (\ref{eq:31}) into (\ref{eq:27.4}), provided that there is no
singularity for  $\mu =1$, the unknown function
$\tilde{\Psi}_{1}(z)$  may be determined:
\begin{equation}
\tilde{\Psi}_{1}(z)=\sqrt{Z_{eff}}(-1/\sqrt{2}+1)\int^{z}_{0}
\frac{\sqrt{2\frac{\partial\Psi_{0}}{\partial z}E
-\beta (\lambda_{0}+B(\frac{\partial\Psi_{0}}{\partial z})^{2})}}
{4\frac{\partial\Psi_{0}}{\partial z}+z\beta B
\frac{\partial\Psi_{0}}{\partial z}+Ez-1}dz
\label{eq:32}
\end{equation}

The expressions (\ref{eq:30}), (\ref{eq:31})  and  (\ref{eq:32})  determine in
the exponential approximation the dependence of the distribution function in
the
exponential region on the energy and the pitch angle  $\mu$.  As
$z\rightarrow 0$,
the obtained solution must coincide with the solution in the polynomial
region for  $y > 1$.  Figure 6 shows both the solutions which are seen to
be almost coincident.

To establish the asymptotical behavior of the
distribution function as  $z\rightarrow\infty$,  we represent the index of the
exponential in (\ref{eq:17}) in the form
\begin{eqnarray}
\label{eq:33}
&\Psi=\Psi_{0}z+\Psi_{1}+\ldots \nonumber &\\
&\Psi_{0}=\varphi_{0}/\delta_{1}+\varphi_{1}/\delta^{1/2}_{1}+
\varphi_{2}+\ldots
&
\end{eqnarray}
$$
z=\epsilon\delta_{1}
$$
Substituting (\ref{eq:33}) into (\ref{eq:7}) with allowance for (\ref{eq:17}),
keeping terms of the order of  $z$ and collecting terms with the same powers
$\delta^{1/2}_{1}$,
we obtain the system of equations for $\varphi_{0},\,\varphi_{1},\,\ldots$:
\begin{eqnarray}
\label{34.1}
&\frac{\partial\varphi_{0}}{\partial\mu}  = 0&\\
\label{eq:34.2}
&\frac{\partial\varphi_{0}}{\partial\mu}\frac{\partial\varphi_{1}}{\partial\mu}
= 0&\\
\label{eq:34.3}
& (\frac{\partial\varphi_{1}}{\partial\mu})^{2}Z_{eff}(1-\mu^{2})-|\mu|\beta \{
\lambda_{0}+B\varphi_{0}^{2} \} + 2\mu\varphi_{0}E = 0 & \\
&2Z_{eff}(1-\mu^{2})\frac{\partial\varphi_{1}}{\partial\mu}
\frac{\partial\varphi_{2}}{\partial\mu} -|\mu|\beta B\varphi_{0}\varphi_{1}+
2\mu E\varphi_{1} +(1-\mu^{2})\frac{\partial\varphi_{1}}{\partial\mu}E=0&
\label{eq:34.4}
\end{eqnarray}
where
$
\beta =\delta_{2}/\delta_{1} , \, \,
$
, and $B$ and $E$  are similar to (\ref{eq:27.3}),  whence
$\varphi_{0}=const$.
Then in the absence
of singularity for $\mu =1$ we have  from  (\ref{eq:34.3}):
\begin{equation}
\label{eq:35} \varphi_{0}=\frac{\sqrt{E^{2}-\beta^{2}\lambda_{0}B}-E} {-\beta
B}
\end{equation}
Knowing $\varphi_{0}$
from  (\ref{eq:34.3}), we obtain
\begin{eqnarray} & \frac{\partial\varphi_{1}}{\partial\mu}=0 & \mu >0
\nonumber
\\
\label{eq:36}
& \frac{\partial\varphi_{1}}{\partial\mu}=-\sqrt{\frac{\mu (4E/\beta B)
(\sqrt{E^{2}-\beta^{2}\lambda_{0}B}-E) } {Z_{eff}(1-\mu^{2})} } & \mu <0
\end{eqnarray}
\begin{eqnarray}
& \varphi_{1}=const& \mu >0 \nonumber \\ &
\label{eq:37}
\varphi_{1}=-\int^{\mu}_{1}\sqrt{\frac{\mu (4E/\beta B)
(\sqrt{E^{2}-\beta^{2}\lambda_{0}B}-E) }
{Z_{eff}(1-\mu^{2})} }d\mu +\tilde{\varphi_{1}}& \mu <0
\end{eqnarray}
where  $ \tilde{\varphi_{1}}$ is an unknown constant.  Then  in the absence of
singularity we obtain from  (\ref{eq:34.4}) for  $\mu = 1$ that
$\tilde{\varphi_{1}}=0$.

The
expressions  (\ref{eq:35}) and  (\ref{eq:37})  determine the asymptotical
behavior of the distribution function for  $E_{e}(r)\neq 0$ as
$z\rightarrow\infty$.  It can be readily seen that in the limit
$z\rightarrow\infty$ (\ref{eq:30})
goes over to  (\ref{eq:35})  and  (\ref{eq:31})  into  (\ref{eq:37}).  So, the
obtained solution  (\ref{eq:30}), (\ref{eq:31})  in the exponential
approximation describes the particle distribution in the entire range of energy
values for  $z>\delta^{1/2}_{1}$.  It should be noted that the asymptotical
expression for the
distribution function for  $z\rightarrow\infty$  (\ref{eq:35}), (\ref{eq:37})
holds also under the condition inverse of  (\ref{eq:31.1}),  but obtaining this
asymptotes is apparently not described by the simple expansion  (\ref{eq:17}).
The solution obtained in this section describes completely the distribution
function of fast particles in the presence of an external field  $E_e$  and
anomalous diffusion in a homogeneous magnetic field.  We have investigated here
a rather general case in the model assumption  $F = const$.  In section 5 we
shall analyze the distributions of fast electrons in specific experimental
conditions, and now we proceed to the question of the influence of
inhomogeneity
of a mean magnetic field.

\section{Distribution function of fast particles in the
    presence of a finite magnetic field gradient.}

We are now ready to consider the influence produced by an inhomogeneous
magnetic
field upon the distribution function of fast electrons.  From the point of view
of physics, the occurrence of terms proportional to the magnetic field gradient
in the kinetic equation (\ref{eq:3}) is associated with keeping the adiabatic
invariant $u_{\perp}^{2}/B(r)=const$  upon a diffusion particle motion along a
discharge.  In this case,  particle diffusion is responsible for particle
energy
redistribution from the transverse degree of freedom to the longitudinal one
and
vice versa, depending on the sign of the derivative  $dB/dr$.  Taking into
consideration the fact that on the average particles move from the center to
periphery of a discharge,  in the case $dB/dr < 0$  part of the transverse
particle energy is transferred into the longitudinal and conversely in the case
$dB/dr > 0$.  Such dynamics must in turn affect the diffusion process itself.
Indeed,  in case the amount of particles with high longitudinal energy
increases
(when  $dB/dr < 0$) or on the contrary decreases  (when  $dB/dr > 0$)  because
the anomalous diffusion coefficient  (\ref{eq:2}) is proportional to the
longitudinal velocity, the diffusion rate will respectively either increase or
on the contrary decrease.

Let us consider the stationary solution of equation
(\ref{eq:3})  in the range  $1\leq\epsilon\leq\delta^{-1/2}$,
where  $\delta =\delta_{1},\: \delta_{2}$.
Not to complicate the picture, we put
the external field  $E_{e}(r)=0$.  In order to obtain the solution, we shall
use the
technique we applied in the solution of equation (\ref{eq:7}) in this range,
that is, the polynomial expansion  (\ref{eq:13}).  The function  $F(r)$
will
put  $F = const$, and $n(r) =  g_0(r)$  of the problem  (\ref{eq:12}).
Figure 7
shows the dependencies of the distribution of fast electrons obtained in the
solution of equation (\ref{eq:7})  depending on the electron energy  $y$  in
two
cases:  $dB/dr > 0$  and $dB/dr < 0$.  The figure also presents for comparison
the result for  $dB/dr = 0$.  As has been expected, the greatest decrement of
$\ln (f/f_{0})$ appears in the case  $dB/dr < 0$, and the smaller in the case
$dB/dr > 0$.  It would also be of interest to trace the behavior of the
solution obtained depending on the longitudinal
$u^{2}_{\|}$ and the  transverse  $u^{2}_{\perp}$
energies.  In Fig.8 we can see the dependencies of
$\ln (f/f_{0})$  on the
transverse and the longitudinal particle energy in the cases  $dB/dr < 0$,
$dB/dr > 0$  and  $dB/dr = 0$.
We see that both for $dB/dr = 0$ and  $dB/dr > 0$
there exists anisotropy in particle distribution over transverse and
longitudinal energy, so that  $T_{\perp }>T_{\| }$.
However  for  $dB/dr < 0$  the anisotropy changes
sign and then  $T_{\perp }<T_{\| }$,
which is naturally connected with a substantial energy
redistribution from the transverse degree of freedom into the longitudinal one.
This effect is most clearly pronounced in the dependence of the distribution
function on the pitch-angle  $\mu$,  Fig.9,  which differs substantially from
that obtained in the case  $dB/dr = 0$,  Fig.1.
The fall in the center  (for $\mu\sim 0$)
in Fig. 9  and the appearance of maxima near  $\mu\sim  1$  is just due to
particle efflux from the region  $\mu\sim 0$  towards the region  $\mu\sim 1$.

\section{Behavior of the distribution function of fast
     electrons in  RFP.}

In the preceding sections of the paper we
have analyzed the influence of various
factors upon the distribution
function of fast particles.  It should be noted here that just in  RFP
discharges the contribution from each of them will be substantial
because all of them are present in this type of devices.  An  RFP
discharge possesses a force-free
magnetic configuration with components of the mean magnetic field
(the toroidal
$B_{z}(r)$ and poloidal  $B_{\theta}(r)$)  which are of the same order of
magnitude and are determined by one and the same parameter
$\Theta =2\pi a I/\Phi$,  where  $I$  is
total current,  $a$  is small radius of the torus,  and  $\Phi$  is the total
toroidal magnetic field flux.  Magnetic field components are well
enough described by Bessel functions of the form $$B_{z}(r)=B_{0}J_{0}(2\Theta
r)
$$

\begin{equation}\label{bes}
\end{equation}
     $$B_{\Theta}(r)=B_{0}J_{1}(2\Theta
r)$$
see Fig.10,  \cite{bodin}.  So, the scale of the
magnetic field gradient in  RFP  will be of the order of the small radius  $a$
of the torus :
$$\frac{dB}{dr}\frac{a}{B}\sim 1$$

Since plasma  in  RFP  is heated in an ohmic way, present in
the discharge is a sufficiently strong longitudinal electric field  $E_e$.
Since an applied  electric field  $\vec {E}=E \vec{e_{z}}$  is
toroidal, its projection  $E_{e}(r)$  onto a magnetic
field line is described by the relation \begin{equation}  \label{eq:38}
E_{e}(r)=E\frac{B_{z}(r)}{B(r)} \end{equation}

\noindent
{}From (\ref{eq:38}) it becomes clear that the applied field is strongly
inhomogeneous about the space.

Direct measurements of distribution of
superthermal electrons in  steady - state of RFP  were carried out on  ZT-40M
devices \cite{16,eks2,eks3,eks4}.  For comparison with the theory, we shall
consider the experimental data obtained in  \cite{eks2}.  Performed in the
experiment were direct measurements of energy distribution of fast electrons in
the near-boundary region of discharge with a temperature  $T_{w}\simeq 20 eV$
near the boundary and $T_{0}\simeq 220 eV$ in the center of discharge.  The
characteristic value of the parameter $\Theta =1.4$.
Observations were carried out for particles of energy up to  $1.5 keV$.
Against
the background of cold plasma with temperature  $T_{w}\simeq 20 eV$ one could
see a tail of energetic electrons which moving
in the direction of the magnetic field and have Maxwell distribution with
characteristic temperature $T_{1}\simeq 530 eV$
and a flux  of particles that moved backward and had a temperature  $T_2
\simeq 330 eV$. The
anisotropy  in the distribution of particles in the longitudinal and
transverse energies was observed, the temperature being  $T_{\perp }<T_{\|}$.
Under conditions of the experiment described above, the parameters
$\delta_1,\:\delta_2$ of our problem were  $\delta_1 = 0.21,  \delta_2 = 0.036$
for $T_{0}=220\, eV,\: \: n=2,4\cdot 10^{13}cm^{-3},\: \: E_{e}=10\, V/m$
and the amplitude of fluctuations  $| b/B | \sim  1 \% ,\;
\frac{F_{max}}{B_0^2a}\sim 4\cdot 10^{-4}$.

The solution we obtained in section three (formulae
(\ref{eq:30}), (\ref{eq:31}), (\ref{eq:32})) gives qualitative agreement
with experiment: exponential distribution of
particles on energy and anisotropy
in particles distribution over forward and backward direction along
magnetic field lines.

An important comparison with experimental data is given by modeling
superthermal electron current $I_{eea}$,
collected by the electrostatic energy
analyzer \\ (EEA), as function of applied retarding potential $V$.
The current is related to the electron distribution function $f(r,\mu
,\epsilon)$ by expression:
\begin{equation}\label{cur}
I_{eea}(r,V)\propto \int^{\infty }_{v_0}\int^{\infty }_{0}f(r,\mu ,\epsilon )
v_{\perp }dv_{\perp }v_{\parallel }dv_{\parallel }
\end{equation}
where
$$
v_0=\sqrt{2e\, V/m_e}
$$

Figure 11 shows the
dependencies of current $I_{eea}$ calculated by formulae
(\ref{eq:30}), (\ref{eq:31}), (\ref{eq:32}), (\ref{cur}) for the given values
of
the parameters $\delta_1$  and  $\delta_2$ for particles flying along the
direction of the magnetic field and backward. The profile of the correlation
function  $F(r)$  shown in Fig.12 was borrowed from ref.\cite{eks2}.
The graph of the zero eigenfunction of the problem
(\ref{eq:12}) $g_{0}(r)$  for a given  $F(r)$  is shown in Fig.13, where
$\lambda_0 = 1.2$.  The profile  $n(r)$  was chosen in the form $n(r) =
g_0(r)$ and $T(r)=1-r^2$.  The calculated current $I_{eea}$ was normalized on
experimental value. The absolute value of $I_{eea}$ seems to be rather
uncertain
because of a lack of information on the correlation function near the
boundary.

Figure 11 testifies to not only
qualitative agreement between the theory and experiment  (Maxwell
distribution and the anisotropy observed in the distribution of particles
flying
in the direction of the magnetic field and backward), but to quantitative one
as
well.  So, the characteristic temperature $T^{teor}_{1}$  for particles moving
along the magnetic field makes up $T^{teor}_{1}=526 \: eV$.  But for particles
moving in the backward direction it was $T^{teor}_{2}=183 \: eV$, i.e. less
than
experimental one.  Backward flowing component of fast electrons  is seen to be
a
much smaller than forward one (less than 10 \% ).  This value is also close to
experimental observations \cite{eks2}.

The behavior of distribution of fast
electrons on longitudinal and transverse energy is shown on fig.14, at the
parameters mentioned above. In this case for particles moving
along the magnetic field line  $T_{\perp}<T_{\|}$ and $T_{\|}=513 eV,\:
T_{\perp}=385 eV$.

\section{Summary}

In conclusion we shall present the main results obtained in this paper.  In the
entire energy range above thermal energies,  we have constructed the solution
of
the kinetic equation describing the distribution function of fast particles in
the presence of an external electric field, collisions, and anomalous diffusion
due to magnetic fluctuations.  The ambipolar electric field resulting from the
difference in the ion and electron diffusion rates is shown to play essential
role in the process of diffusion and to be responsible for the fact that the
character of the diffusion depends strongly on the density profile of the main
plasma particles and on the profile of an external electric field applied to
the
plasma.  The influence of inhomogeneity of the magnetic field of a discharge
upon the distribution function of fast particles is investigated. The analysis
performed suggests that the magnetic field inhomogeneity is an important factor
affecting  the distribution function of fast particles.  In the polynomial
region of energy values, which is an immediate neighbor of the thermal region,
the semi - analytical solution is
constructed with allowance for all the factors
mentioned above.  On the basis of comparison with experimental data we have
shown that the theory describes  the basic details  of the distribution
function
of fast particles. We have referred our analysis mainly to  the case of RFP
discharges, but the results obtained in the absence of strong electric field
seems to be closely connected with anomalous process in TOKAMAK discharges.

\section*{Acknowledgments}
The authors wish to thank V.L.Ginzburg and Richard Nebel
for stimulating discussions related to this work. We would
like also to acknowledge the referees for the comments, which
undoubtedly promote for improvement of the paper.

This work was performed under the subcontract Y6010 001 4-9U with LANL.

One of the authors (AL) would like to thank Landau Scholarships
for financial support.

\clearpage
\section*{Captions}

\noindent
Fig.1 The dependencies of fast particles distribution function on pitch angle
$\mu$ into polynomial area of solution
under various parameters $\beta$,
$y=2,\; r=0.5$. Dashed line - solution, obtained in \protect\cite{gr1} under
$\beta=0$.

Fig.2 The dependencies of fast particles distribution function
$\ln (f/f_0)$ averaged over $\mu$ on energy $y$ into polynomial solution area.
1) - $n(r)=g_0(r)$, $T(r)=const$, $E_e(r)=const$
2) - in case of nonuniform electric field $n(r)=g_0(r)$, $T(r)=const$
3) - $n(r)=\tilde{n}(r)$,
plot  $\tilde{n}(r)$ is shown on figure 3, $T(r)=const$, $E_e(r)=const$
and
4) $E_e(r)=const$, $n(r)=g_0(r)$, $T(r)=1-r^2$.
$\beta =0.3, \; r=0.5$.

Fig.3 The distribution over discharge of initial density profile
$\tilde{n}(r)$, distribution function $f(r)$ at $y=1$, obtained in solving
the system (\protect\ref{eq:16})  and  eigenfunction of the problem
(\protect\ref{eq:12}) (dashed line).

Fig.4 The solution of equation  (\protect\ref{eq:20})  as a function of $\xi$.

Fig.5 The fast electrons distribution function $f/f_0$  as a function of
$\mu$ in the absence of applied electric field  $E_e$. Solid line - the
polynomial solution
 at $y>1$, dashed line - the exponential solution
 at $z=y\delta_2^{1/2}\ll 1$.

Fig.6 The fast electrons distribution function $f/f_0$  as a function of
$\mu$ with the applied electric field $E_e$ present. Solid line - the
polynomial solution
 at $y>1$, dashed line - the exponential solution
 at $z=y\delta_2^{1/2}\ll 1$.

Fig.7 The dependencies of fast electrons distribution function, averaged over
$\mu$, $\ln (f/f_0)$ on energy $y$ with finite gradient of mean magnetic field
present.  1) $dB/dr>0$, 2) $dB/dr=0$, 3) $dB/dr<0$. $r=0.95$

Fig.8 The dependencies of fast electrons distribution function $\ln (f/f_0)$ on
longitudinal (solid line) and transverse (dashed line) energy, in the present
of
finite gradient of mean magnetic field.  1) $dB/dr>0$, 2) $dB/dr=0$,
3) $dB/dr<0$. $r=0.95$

Fig.9 The dependencies of fast electrons distribution $ f/f_0$
on $\mu$ under various positions over radius. $dB/dr<0$.

Fig.10 Radial distribution of magnetic field components across
the minor radius.

Fig.11 The comparison of theoretical versus experimental  current
$I_{eea}$, collected by electron energy analyzer as a function of
retarding potential $V$ for the case of ZT-40M. Solid line - for particles
moving along the magnetic field line, dashed line - for particles moving
backward.  Experimental data  are shown by labels.

Fig.12 Correlation function of fluctuation in RFP.

Fig.13 Eigenfunction of the problem (\protect\ref{eq:12})
for the correlation function, depicted on figure 12.

Fig.14 The dependencies of fast electrons distribution function $\ln (f)$ on
longitudinal (solid line) and transverse (dashed line) energy for particles
moving along the magnetic field line. $\beta =0.17$, $\delta_{1}=0.21,\:
\delta_{2}=0.036$.

\end{document}